\documentclass[journal]{IEEEtran}
\usepackage{amsmath,amsfonts}
\usepackage{algorithmic}
\usepackage{algorithm}
\usepackage{array}
\usepackage[caption=false,font=normalsize,labelfont=sf,textfont=sf]{subfig}
\usepackage{textcomp}
\usepackage{stfloats}
\usepackage{url}
\usepackage{verbatim}
\usepackage{graphicx}
\usepackage{cite}
\usepackage{subfig}
\usepackage{balance} 
\usepackage{booktabs}

\begin{document}

\title{DeepRT Engine: A Unified GPU-Parallel Ray-Tracing Framework with Hybrid SBR-IM Path Search for 6G Digital Twin Channel}

\author{
	Tao~Wu,~\IEEEmembership{Student Member,~IEEE},
	Li~Yu,~\IEEEmembership{Member,~IEEE},
	Yuxiang~Zhang,~\IEEEmembership{Member,~IEEE},
	Jianhua~Zhang,~\IEEEmembership{Fellow,~IEEE},
	Qixing~Wang,~\IEEEmembership{Member,~IEEE},
	and~Guangyi~Liu,~\IEEEmembership{Member,~IEEE}
	\thanks{This work was supported in part by the National Natural
		Science Foundation of China under Grants 62401084 and 62525101,
		in part by the National Key Research and Development Program
		of China under Grant 2023YFB2904805, and in part by the Beijing
		University of Posts and Telecommunications--China Mobile
		Communications Group Co., Ltd. Joint Institute.}
	\thanks{Tao Wu, Li Yu, Yuxiang Zhang, and Jianhua Zhang are with
		the State Key Laboratory of Networking and Switching Technology,
		Beijing University of Posts and Telecommunications,
		Beijing 100876, China
		(e-mail: \{vvt, li.yu, zhangyx, jhzhang\}@bupt.edu.cn).
		Li Yu is the corresponding author.}
	\thanks{Qixing Wang and Guangyi Liu are with
		China Mobile Research Institute,
		Beijing 100053, China
		(e-mail: \{wangqixing, liuguangyi\}@chinamobile.com).}
}

\markboth{Journal of \LaTeX\ Class Files,~Vol.~14, No.~8, August~2021}%
{Shell \MakeLowercase{\textit{et al.}}: A Sample Article Using IEEEtran.cls for IEEE Journals}


\maketitle

\begin{abstract}
Digital twin channel (DTC) aims to establish a real-time digital counterpart of physical wireless channels for reproducing and predicting site-specific propagation characteristics. As a high-precision channel computation method for realistic propagation scenarios, ray tracing (RT) serves as a key enabler for DTC construction. However, conventional RT suffers from high complexity under serial path-searching workflows. This letter proposes DeepRT Engine (DeepRT-E), a parallel RT acceleration architecture with a three-stage physically-inspired pipeline for real-time DTC construction. Firstly, DeepRT-E constructs a bounding volume hierarchy (BVH) to partition the scene and reduce redundant ray-surface intersections. Secondly, the shooting and bouncing rays (SBR) algorithm is executed through a ray-level parallel tracing framework to identify candidate surface sequences and prune the search space of the image method (IM). Finally, a parallel batched IM solver refines the retained candidates for accurate propagation-path recovery. Simulation results show that DeepRT-E reduces runtime by 96.3$\%$ and achieves a converged error of only 0.001 dB, outperforming Wireless InSite and Sionna in efficiency and accuracy.
\end{abstract}


\section{Introduction}

\IEEEPARstart{T}{he} evolution toward 6G environment intelligence communication requires accurate, environment-consistent, and low-latency channel data for beam management, resource allocation, and network optimization~\cite{ref_engineering}. Digital twin channel (DTC) refers to the real-time digital mapping of physical wireless channels~\cite{ref_DTC}, where radio environment information is utilized to reproduce and predict site-specific channel characteristics~\cite{ref_REKP}. Such DTC-oriented channel data are essential for 6G air-interface design and channel foundation model training~\cite{ref_6G,ref_ChannelGPT}. Ray tracing (RT) provides deterministic channel prediction from environmental geometry and electromagnetic properties, making it effective for DTC construction~\cite{ref_rt,ref_thz_rt}. Our previous DeepRT work uses RT-derived channel data as physical priors for artificial intelligence (AI)-based DTC generation, motivating an efficient RT engine to provide low-latency and physics-consistent ray-level data~\cite{ref_DeepRT}. However, conventional RT still suffers from high path-searching complexity in complex scenarios, especially for high-order reflections.

Existing RT platforms have significantly promoted site-specific wireless channel modeling. Wireless InSite provides mature engineering-level RT capabilities and has been widely used for deterministic propagation simulation~\cite{ref_WI}. Sionna supports graphics processing unit (GPU)-accelerated and differentiable RT, making RT more accessible for learning-based communication systems~\cite{ref_Sionna}. Nevertheless, for real-time-oriented DTC construction, there remains an efficiency--accuracy tradeoff among massive ray launching, candidate-path pruning, exact path recovery, and repeated channel updates. Therefore, an RT engine that jointly considers path-searching accuracy, candidate-space compression, and GPU-parallel execution is still required.

To improve the speed and accuracy of RT-based path search~\cite{ref_effi}, various enhanced algorithms have been proposed. In~\cite{ref_r1}, shooting and bouncing rays (SBR) is used for coarse path identification and the image method (IM) is subsequently applied for exact path recovery in suburban channel analysis. GPU-based parallel SBR implementations have been developed to accelerate massive ray launching and ray-surface intersection tests~\cite{ref_r2,ref_r3}, while parallel IM acceleration has also been investigated~\cite{ref_r5}. However, existing studies mainly focus on either isolated acceleration of SBR/IM or algorithm-level SBR-IM combination. An integrated RT engine that jointly supports hybrid path search, GPU-parallel execution, and DTC-oriented channel generation remains insufficiently explored.

This letter proposes DeepRT Engine (DeepRT-E), a parallel RT acceleration framework for real-time DTC construction. DeepRT-E integrates bounding volume hierarchy (BVH)-based scene representation, SBR-IM hybrid path search, and channel data generation into a unified physics-inspired computation engine for DTC-oriented applications. Specifically, DeepRT-E decomposes the hybrid path-search workflow according to the physical structure of ray propagation and maps it onto a two-level parallel architecture. BVH-based spatial partitioning accelerates ray–surface intersection tests through hierarchical scene traversal, while SBR candidates exploration is executed concurrently across independent rays. The retained candidate sequences are then processed by batched IM kernels for parallel exact path recovery. Compared with Sionna’s general-purpose parallel RT framework, DeepRT-E adopts a hardware-oriented kernel architecture that exploits fine-grained parallelism to enable real-time channel generation for DTC.


\section{DeepRT-E Framework}
\subsection{Architecture of DeepRT-E}

\begin{figure}[h]
	\centering
	\includegraphics[width=\linewidth]{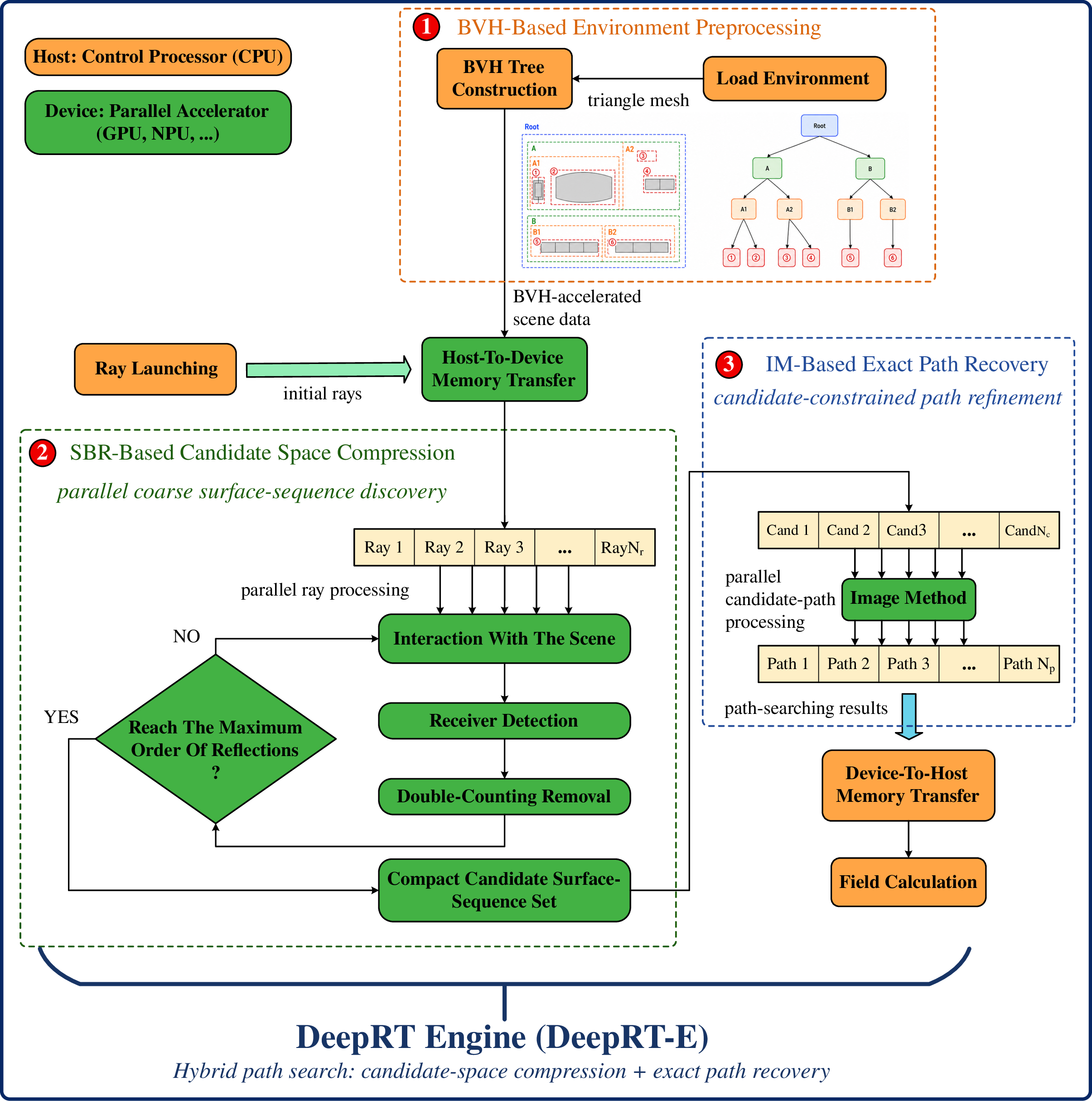}
	\caption{Overall architecture of DeepRT-E for real-time DTC construction.}
	\label{parallel_flow}
\end{figure}

Fig.~\ref{parallel_flow} illustrates the overall architecture of the proposed DeepRT-E framework. DeepRT-E integrates BVH-based environment preprocessing, SBR-based candidate space compression, IM-based exact path recovery, and channel data output in a pipeline-oriented manner. 
First, in the BVH-based environment preprocessing stage, the environment is loaded on the host side, where a BVH is constructed over the simulation scene to reduce redundant ray-surface intersection tests. The initial rays are generated by subdividing an icosahedron according to a specified subdivision level, ensuring uniform angular coverage. The generated ray set and BVH-accelerated scene data are then transferred from the host to the GPU for parallel processing.

Second, in the SBR-based candidate space compression stage, DeepRT-E performs parallel coarse surface-sequence discovery on the device. Multiple rays are traced concurrently, and the interacted surface-index sequence is recorded when a ray is received. After double-counting removal, repeated sequences generated by different received rays are eliminated, producing a compact candidate surface-sequence set for subsequent IM solving. This avoids storing intermediate reflection-point information during SBR and reduces the search space for accurate path recovery.

Third, in the IM-based exact path recovery stage, the compressed candidate surface sequences are further processed by the IM module. Different from conventional IM, which requires exhaustive enumeration of possible surface combinations, DeepRT-E applies IM only to the SBR-compressed candidate sequences. The IM module evaluates these candidates in parallel and recovers accurate propagation paths with exact reflection points and path parameters. Based on the recovered propagation paths, field calculation is then performed to generate channel results for real-time DTC construction.

\subsection{SBR-IM Hybrid Path Search}

\begin{figure}[h]
	\centering
	\subfloat[]{\includegraphics[width=0.46\linewidth]{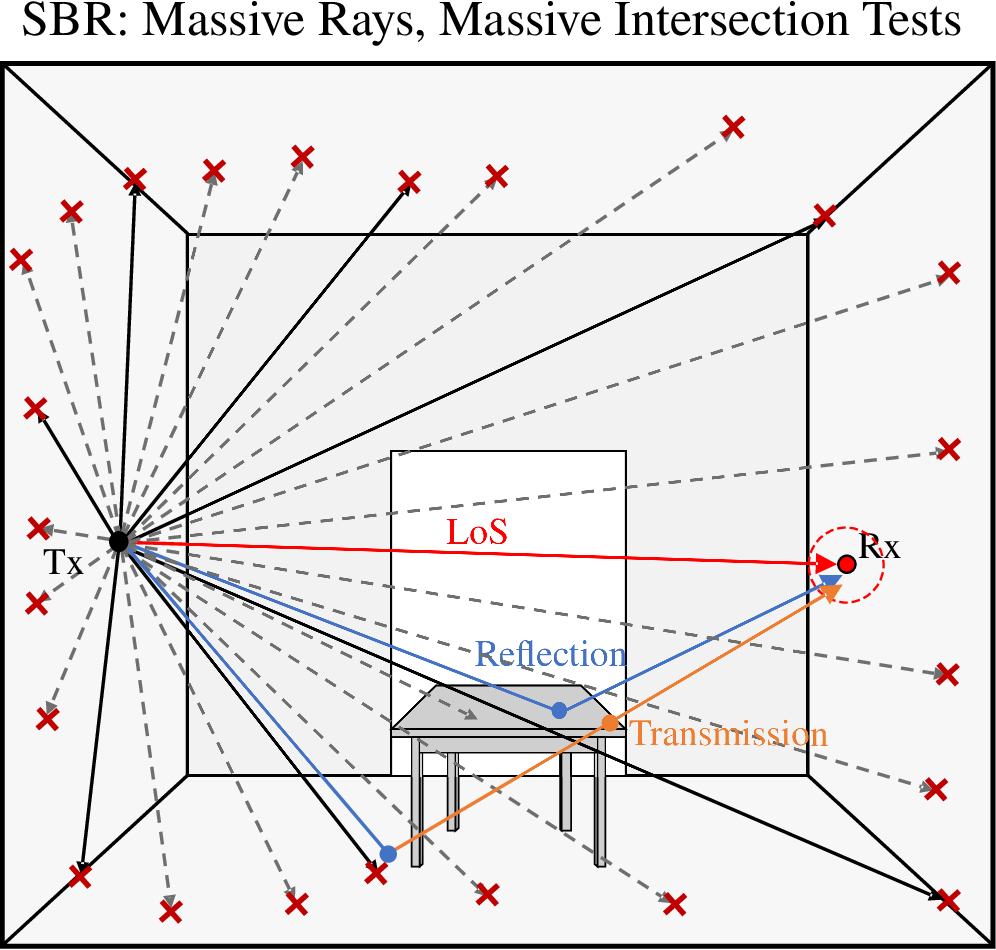}%
		\label{fig_SBR}}
	\hspace{0.1in}
	\subfloat[]{\includegraphics[width=0.46\linewidth]{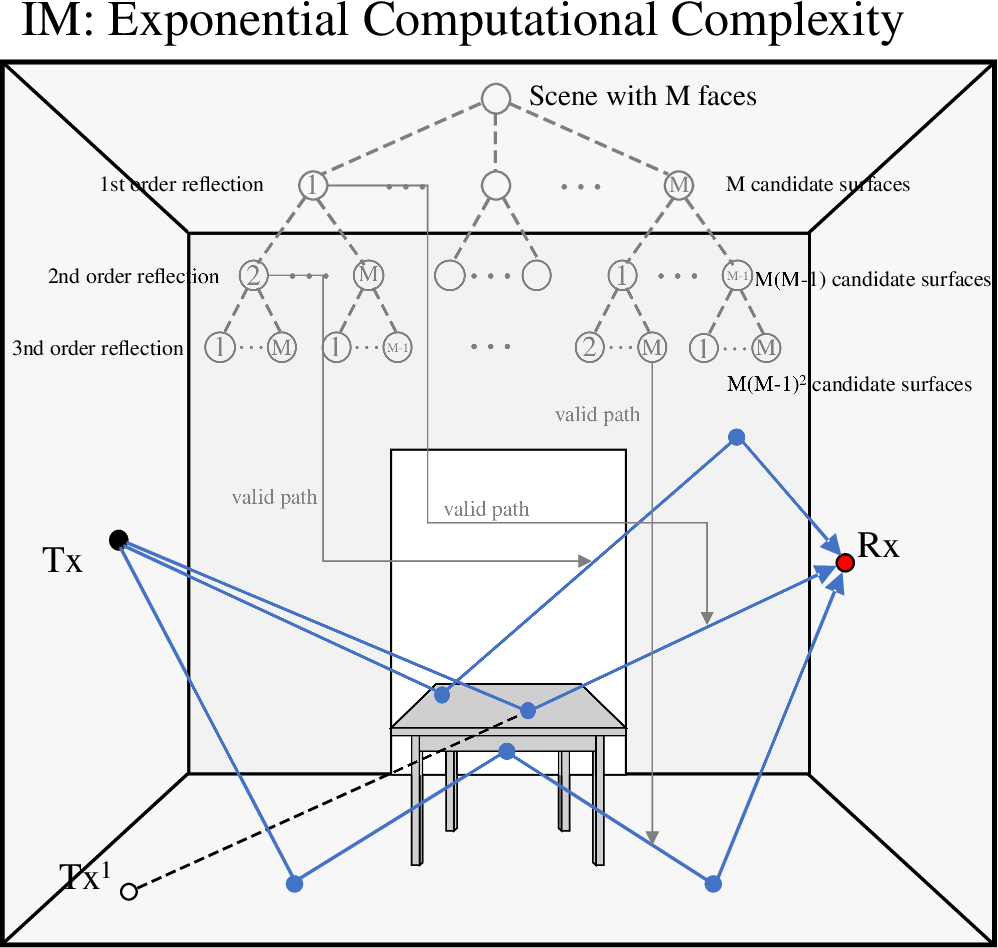}%
		\label{fig_IT}}
	\caption{Comparison of SBR and IM path-searching mechanisms, (a) shooting and bouncing rays, (b) image method.}
	\label{fig_SBR_IT}
\end{figure}

The second stage of DeepRT-E is the hybrid SBR-IM path search, which combines forward candidate discovery and exact path recovery. As shown in Fig.~\ref{fig_SBR_IT}\subref{fig_SBR}, the SBR method launches multiple rays from the transmitter and traces their interactions with the scene until the rays either reach the receiver region or exit the scene boundary. Initial rays are generated through uniform sampling of points on a sphere centered at the transmitter~\cite{SBR_sampling}. Common sampling techniques include normalized random distribution, Fibonacci spiral sampling, and Platonic solid subdivision. In this work, we adopt a subdivided icosahedron for sampling, where each triangular face of the icosahedron is divided into several meshes and the rays are launched toward the vertices of these subdivided meshes.

During SBR tracing, ray-surface intersection tests are performed iteratively to detect interactions with the scene geometry. In conventional SBR, the interacted surfaces and the corresponding intersection coordinates can be recorded during ray tracing. In DeepRT-E, only the interacted surface-index sequence is retained for subsequent IM-based path recovery, since the exact interaction points and path parameters are finally determined by the IM module. At the receiver, a reception sphere with radius $r$ is established, and a ray is considered received once it intersects this sphere. The reception radius is calculated as \(r=\alpha d/\sqrt{3}\)~\cite{SBR1}, where $\alpha=\tan^{-1}\left(\frac{12}{(n-1)\sqrt{3}(3+\sqrt{5})}\right)$ is determined by the icosahedral subdivision level $n$, and $d$ denotes the total propagation distance. Since multiple rays within a certain angular range may intersect the reception sphere, identical surface-index sequences can be recorded repeatedly, resulting in double counting. Therefore, duplicate sequences are removed to obtain a compact candidate surface-sequence set.

IM employs the method of images to identify specular reflection paths, as illustrated in Fig.~\ref{fig_SBR_IT}\subref{fig_IT}. Based on a preset reflection order $N$, conventional IM enumerates possible surface combinations in the scene and verifies whether valid reflection paths exist between the transmitter and receiver. IM can accurately recover reflection points and path parameters, but its computational complexity increases rapidly with the reflection order. In contrast, SBR avoids exhaustive enumeration, but its accuracy depends on ray density and receiver-detection settings. A sparse ray distribution improves efficiency but may miss valid paths, whereas a denser distribution improves accuracy at the expense of increased ray-surface intersection costs.

DeepRT-E exploits the complementary characteristics of SBR and IM. The SBR stage is used for coarse candidate-sequence discovery, while the IM stage is applied only to the retained candidate surface sequences for exact path recovery. Therefore, DeepRT-E avoids exhaustive IM enumeration over all surface combinations and reduces the candidate space before accurate path computation. By integrating SBR-based candidate compression and IM-based path recovery, DeepRT-E improves the efficiency--accuracy tradeoff in complex propagation environments with high-order reflections.

\section{Parallel Execution Architecture}

\begin{figure}[h]
	\centering
	\includegraphics[width=\linewidth]{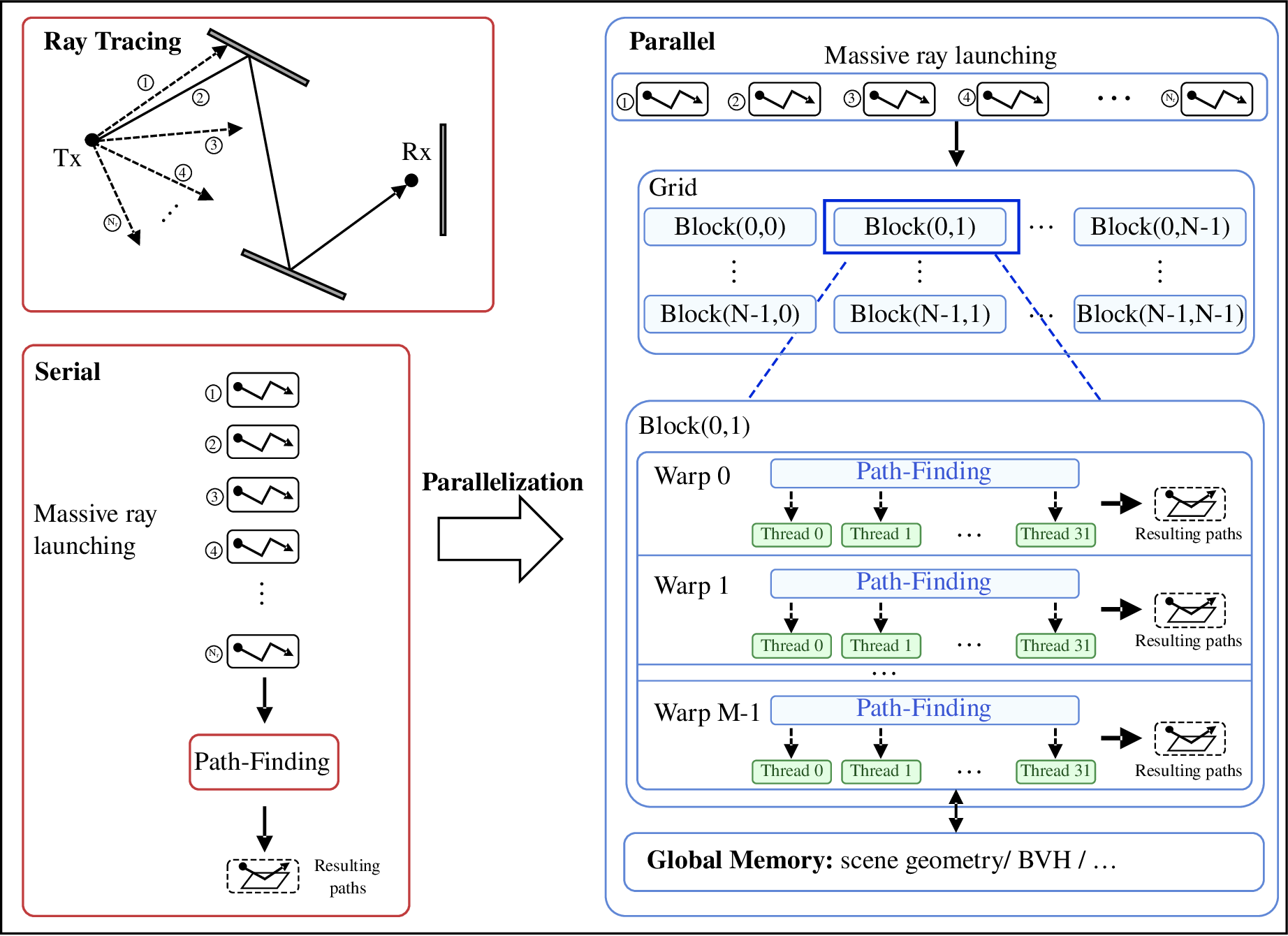}
	\caption{Parallel execution architecture of DeepRT-E.}
	\label{parallel}
\end{figure}

In the process of ray tracing for multipath identification, the computation of each individual ray is independent of the others. Therefore, a parallel architecture can be adopted to process a large number of rays simultaneously. As illustrated in Fig.~\ref{parallel}, the parallel execution is organized through a grid--block--warp--thread hierarchy, where each thread traces an individual ray or candidate path, and threads within the same warp execute the same path-searching kernel on different data. In this way, the conventional ray-by-ray loop is mapped to concurrent thread-level execution on GPU, thereby significantly improving the efficiency of the path-searching process.

For a scene consisting of $M$ surfaces and a predefined maximum reflection order $N$, the computational complexity of the conventional SBR method is $\mathcal{O}\left(MNx\right)$, where $x$ denotes the number of rays launched from the transmitter. In contrast, directly applying the IM method results in a complexity of $\mathcal{O}\left(M^N\right)$, which becomes extremely time-consuming for high-order reflections in complex environments. Therefore, path search becomes the major bottleneck in conventional RT, which motivates parallelization. In DeepRT-E, BVH reduces the average intersection cost from $\mathcal{O}(M)$ to approximately $\mathcal{O}(\log M)$, and IM is applied only to the compact candidate set $\mathcal{C}_{\mathrm{SBR}}$ obtained after reception detection and double-counting removal. Therefore, the path-searching complexity of DeepRT-E can be approximated as
\begin{equation}
	\label{eq:deeprt_complexity}
	T_{\mathrm{DeepRT\text{-}E}}
	=
	\mathcal{O}
	\left(
	xN\log M
	+
	\left|
	\mathcal{C}_{\mathrm{SBR}}
	\right|
	\right).
\end{equation}
Here, the two terms correspond to BVH-accelerated SBR candidate discovery and candidate-constrained IM path recovery, respectively.

\section{Simulation and Analysis}
\begin{figure}[htbp]
	\centering
	\begin{minipage}{0.48\linewidth}
		\centering
		\subfloat[]{\includegraphics[width=0.8\linewidth]{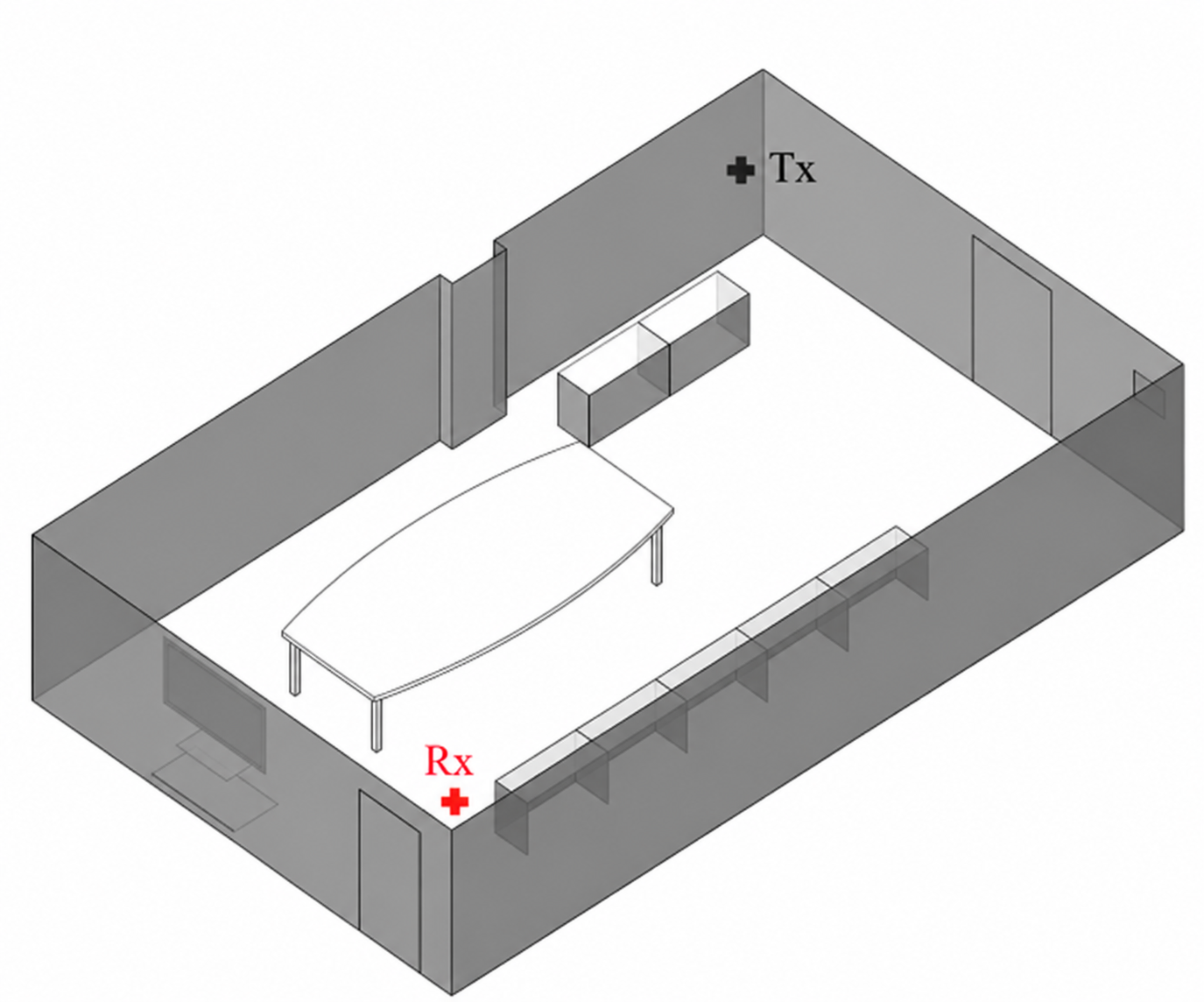}
			\label{indoor_scene}}
	\end{minipage}\hfill
	\begin{minipage}{0.48\linewidth}
		\centering
		\subfloat[]{\includegraphics[width=0.8\linewidth]{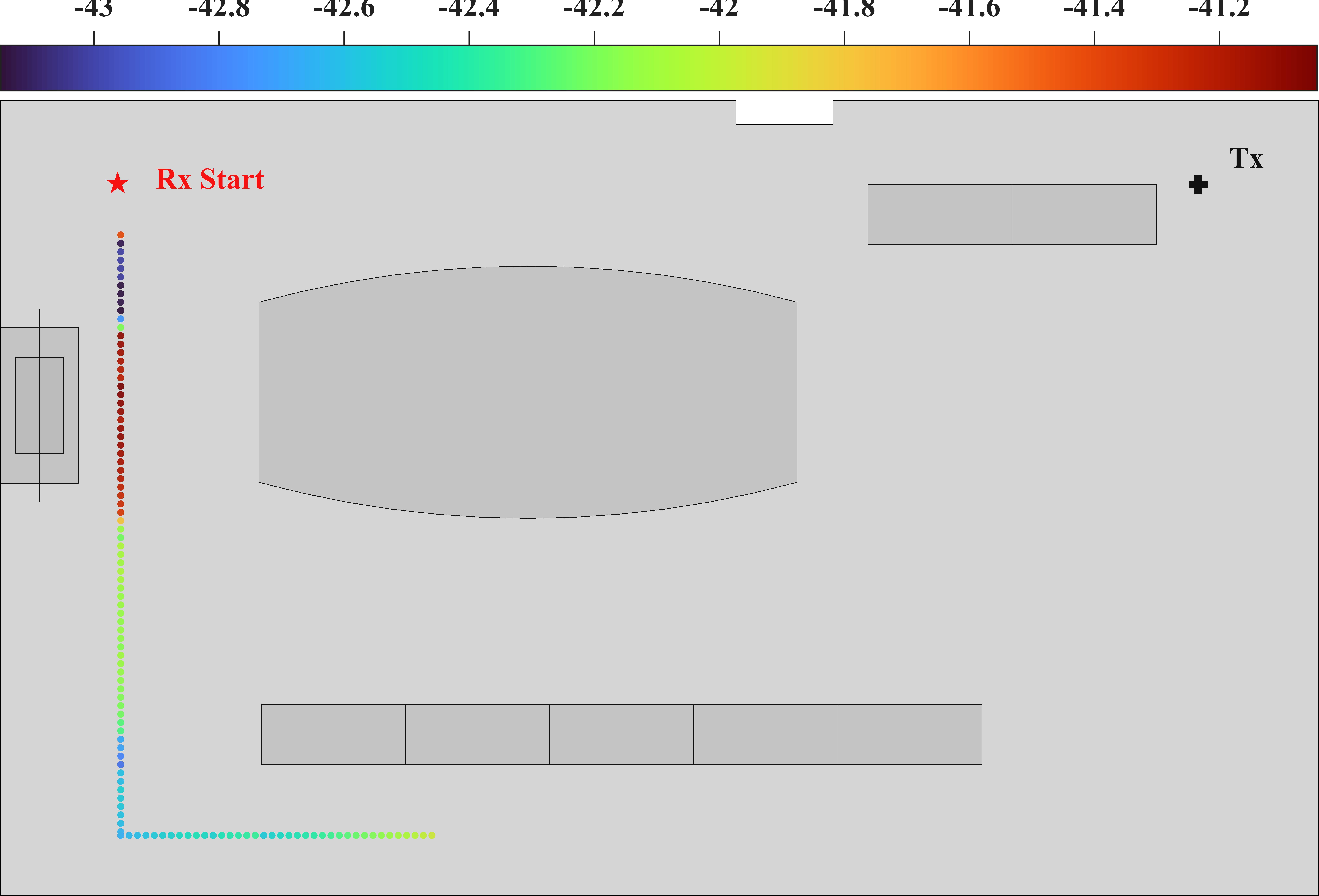}
			\label{indoor_route}}
	\end{minipage}
	
	\caption{Simulation scenario for performance evaluation: (a) the indoor geometric model; (b) the receiver movement trajectory.}
	\label{scene}
\end{figure}

\begin{figure}[h]
	\centering
	\includegraphics[width=0.7\linewidth]{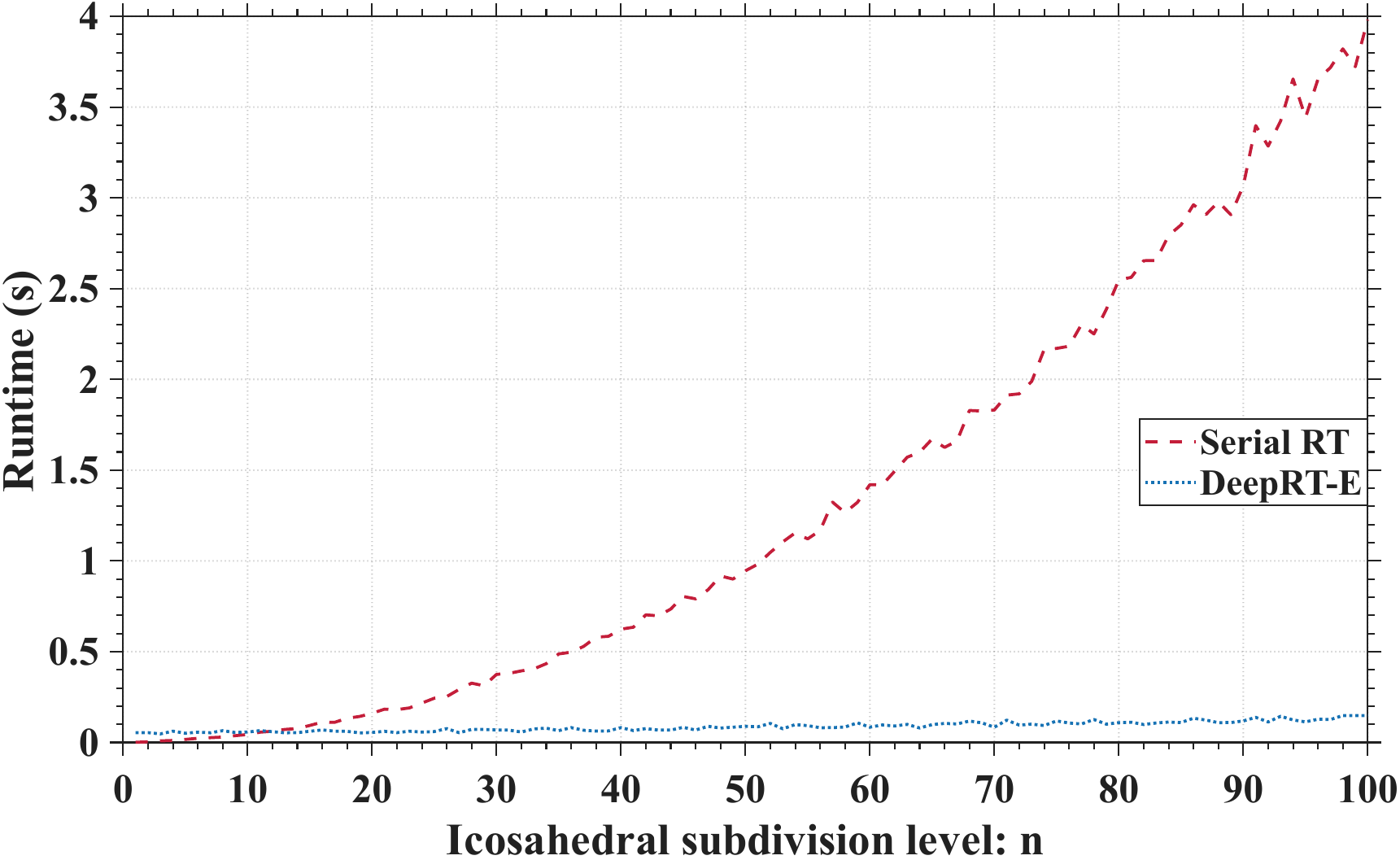}
	\caption{Runtime comparison between serial RT and DeepRT-E.}
	\label{time_compare}
\end{figure}

To validate the efficiency of the parallel algorithm, the execution times before and after parallelization were compared in the scenario shown in Fig.~\ref{scene}\subref{indoor_scene}. In the simulations, the number of launched rays was controlled by the subdivision level $n$, and their relationship is formulated as
\begin{equation}
	\label{equ2}
	x=10\left(n-1\right)\left(n-2\right)+30\left(n-1\right)+12,
\end{equation}
where $x$ denotes the total number of launched rays. The comparison results are shown in Fig.~\ref{time_compare}. The simulations were conducted on a server platform equipped with dual Intel Xeon Gold 6330 CPUs (2.00 GHz) and an NVIDIA GeForce RTX 4090 GPU. While the runtime of the conventional algorithm increases as the number of launched rays grows, DeepRT-E maintains high computational efficiency even with a massive ray count, benefiting from its parallel processing architecture. As demonstrated in Table~\ref{table1}, the execution time is reduced by 96.3$\%$ after parallelization.

\begin{figure}[h]
	\centering
	\includegraphics[width=\linewidth]{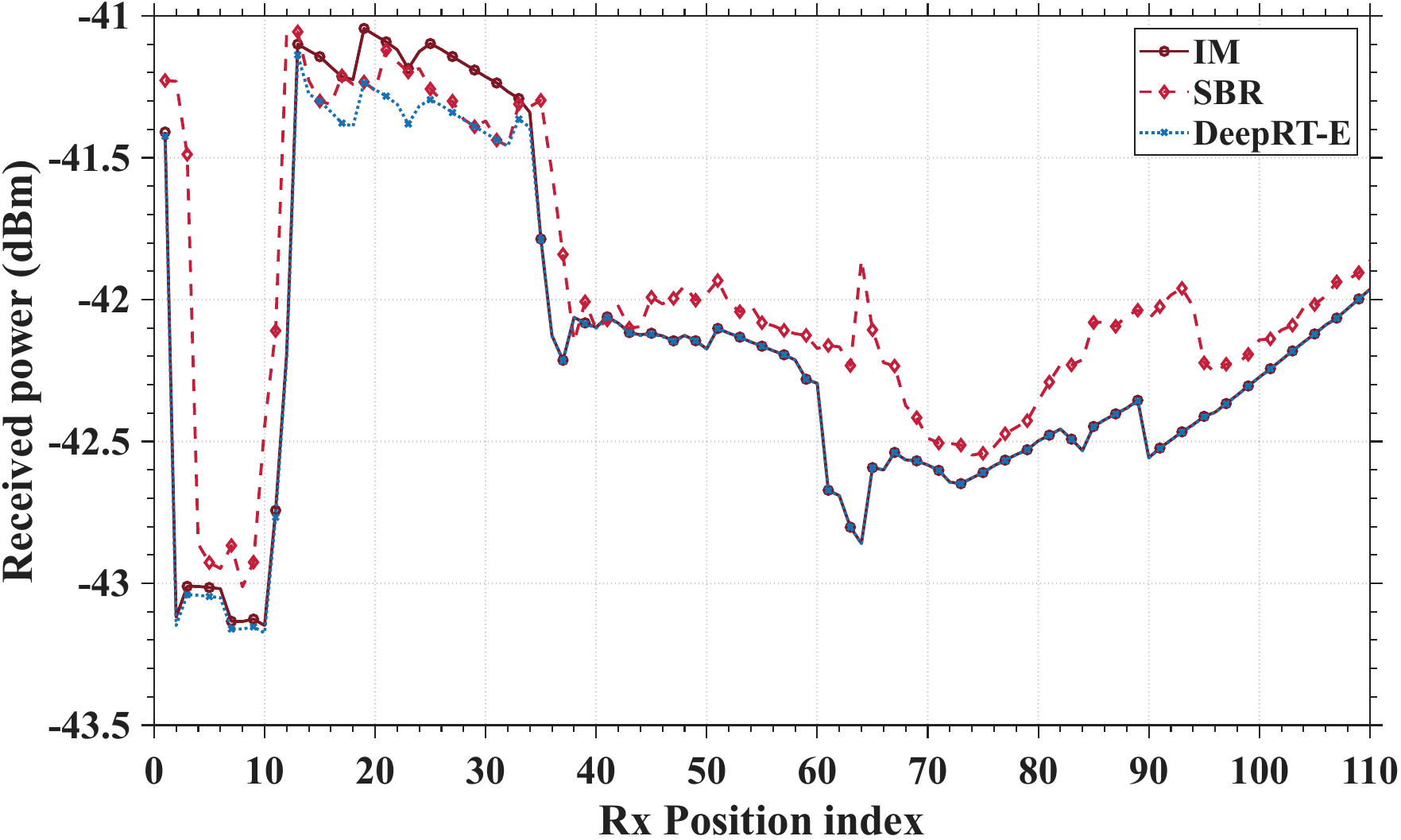}
	\caption{Received power comparison among IM, SBR, and DeepRT-E along the receiver trajectory.}
	\label{SBR_compare}
\end{figure}

The performance gains of the proposed DeepRT-E algorithm over the conventional SBR method were evaluated through comparative simulations conducted along the predefined trajectory depicted in Fig.~\ref{scene}\subref{indoor_route}. During the simulation, the transmitter remained stationary while the receiver moved sequentially along the path. At each receiver location, both ray launching (RL) algorithms configured with $n=100$ were executed, and the received power obtained by DeepRT-E, SBR, and IM along the receiver trajectory is presented in Fig.~\ref{SBR_compare}. It can be observed that the results produced by DeepRT-E closely match those of the IM algorithm, whereas the conventional SBR method exhibits relatively larger deviations. As detailed in Table~\ref{table_error_comparison}, the proposed DeepRT-E method achieves a marginal average error of only 0.02 dB, while the conventional SBR algorithm exhibits a significantly larger average error of 0.14 dB.

For a comprehensive evaluation, a comparative study was conducted in the indoor scenario shown in Fig.~\ref{scene}\subref{indoor_scene} at a carrier frequency of 28 GHz, using ray tracing results obtained from Wireless InSite and Sionna, both of which are based on RL algorithm. Wireless InSite adopts the X3D model, which is built upon the SBR method and employs a reception sphere to detect rays arriving at the receiver. It further applies exact path correction to force rays to terminate at the receiver location, thereby improving the accuracy of power and phase calculations, and is deployed on GPUs for accelerated computation. Sionna employs a hybrid ray-tracing strategy in which initial rays are generated by Fibonacci spiral sampling and the resulting propagation candidates are further processed through an IM-related path validation procedure. Different from the reflection- and transmission-order settings used in DeepRT-E and Wireless InSite, Sionna limits the total number of propagation interactions through a bouncing-depth parameter. Therefore, paths whose total interaction order exceeds the specified bouncing depth were filtered out for fair comparison.

\begin{figure}[h]
	\centering
	\includegraphics[width=\linewidth]{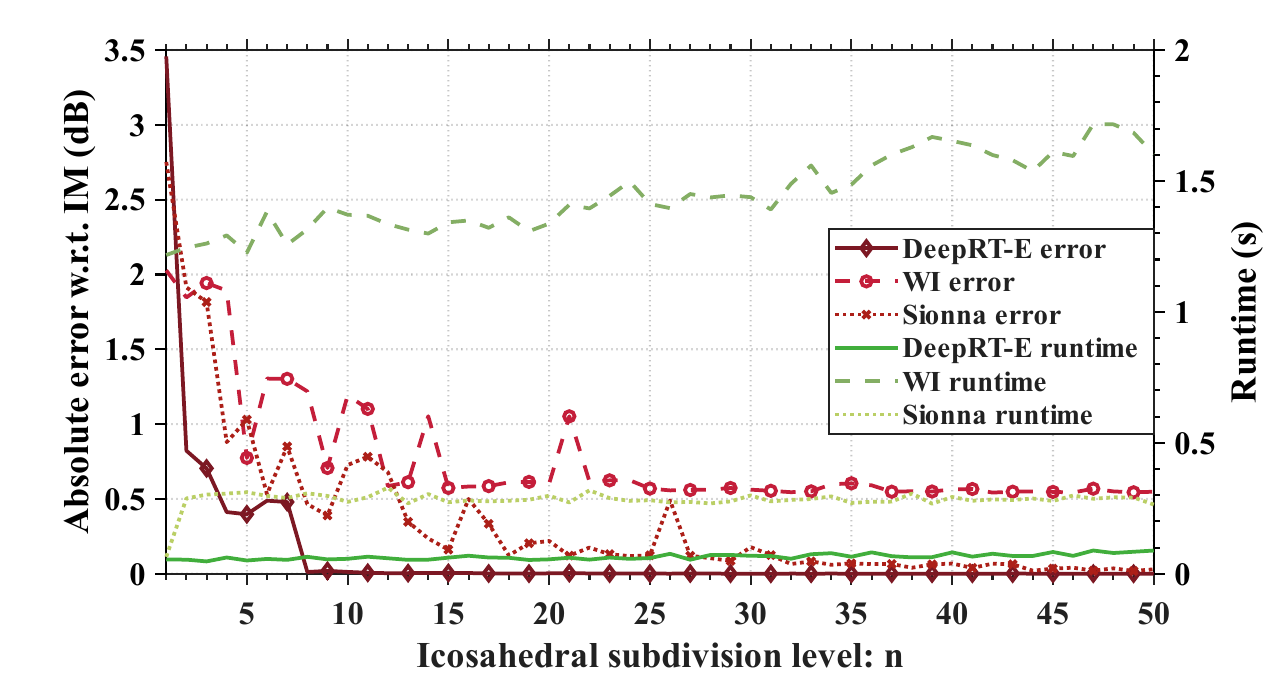}
	\caption{Convergence error and runtime comparison among DeepRT-E, Wireless InSite, and Sionna.}
	\label{sionna_compare}
\end{figure}

Fig.~\ref{sionna_compare} illustrates the simulation error and execution time as a function of the number of launched rays. As the ray count increases, the ray tracing results exhibit a clear convergence trend. Among the evaluated methods, the proposed DeepRT-E exhibits the fastest convergence rate and stabilizes at approximately $n=10$ (corresponding to roughly 1,000 launched rays). In contrast, Sionna converges more slowly and requires approximately $n=30$ to reach a stable state. Regarding accuracy, when benchmarked against the IM algorithm, both DeepRT-E and Sionna achieve exceptionally high precision and exhibit nearly negligible errors after convergence. However, Wireless InSite presents a residual error of approximately 0.55 dB upon convergence, as shown in Table~\ref{table_error_comparison}. In terms of execution time, as the number of rays increases, the simulation times for DeepRT-E and Sionna remain almost constant, demonstrating excellent parallel computing capabilities. Consistent with this observation, the execution time comparison conducted on the same server platform, as summarized in Table~\ref{table1}, indicates that DeepRT-E achieves a substantial speedup over Wireless InSite and demonstrates computational efficiency on the same order of magnitude as Sionna.

\begin{table}[htbp]
	\caption{Runtime comparison of different RT implementations.}
	\label{table1}
	\centering
	\renewcommand{\arraystretch}{1.2}
	\begin{tabular}{ccccc}
		\toprule
		\textbf{Algorithm} & DeepRT-E & serial hybrid RT & Wireless InSite & Sionna\\
		\midrule
		\textbf{Runtime} & 0.148 s & 3.981 s & 2.804 s & 0.286 s \\
		\bottomrule
	\end{tabular}
\end{table}

\begin{table}[htbp]
	\centering
	\caption{Accuracy comparison of different RT algorithms.}
	\label{table_error_comparison}
	\resizebox{\linewidth}{!}{
		\begin{tabular}{ccccc}
			\toprule
			\textbf{Error Type} & \textbf{DeepRT-E} & \textbf{SBR} & \textbf{WI} & \textbf{Sionna} \\
			\midrule
			Average error & 0.021 dB & 0.145 dB & -- & -- \\
			Post-convergence error & 0.001 dB & -- & 0.550 dB & 0.028 dB \\
			\bottomrule
		\end{tabular}
	}
\end{table}

\section{Conclusion}
This letter proposes DeepRT-E to improve the efficiency--accuracy tradeoff in channel generation. By combining SBR-based candidate space compression, IM-based exact path recovery, and parallel execution, DeepRT-E enables efficient and accurate propagation-path computation. Simulation results show that DeepRT-E reduces runtime by 96.3$\%$ compared with conventional serial methods and achieves competitive performance against Sionna and Wireless InSite. These results demonstrate its potential as an efficient physical RT engine for high-fidelity real-time channel generation in DTC construction.
\newpage

\balance

\end{document}